\documentclass[superscriptaddress,pre,twocolumn,showpacs,preprintnumbers,amsmath,amssymb,aps]{revtex4}
\usepackage{graphicx}% Include figure files
\usepackage{epstopdf}
\usepackage{dcolumn}% Align table columns on decimal point
\usepackage{bm}% bold math
\usepackage{color}

\newcommand{\feq}{f^{\rm eq}}

\newcommand{\cs}{c_{\rm s}}

\begin{document}

\preprint{Submitted to Physical Review E}
%\preprint{Submitted to Physical Review Letters}
%\preprint{Submitted to \red{PRE ?}...}

\title{Comment on the paper\\
Li-Shi Luo, Wei Liao, Xingwang Chen, Yan Peng and Wei Zhang, \\
Numerics of the lattice Boltzmann method: Effects of collision models
on the lattice Boltzmann simulations,
Physical Review E 83, 056710 (2011)
}

\author{I.V. Karlin}\email{karlin@lav.mavt.ethz.ch}
\affiliation {Energy Technology Research Group, School of Engineering Sciences, University of Southampton, Southampton, SO17 1BJ, UK}
\affiliation{Aerothermochemistry and Combustion Systems Lab, ETH Zurich, 8092 Zurich, Switzerland}
\author{S. Succi}\email{succi@iac.cnr.it}
\affiliation{Istituto Applicazioni Calcolo CNR, Via dei Taurini 19, 00185 Rome, Italy}
\affiliation{School of Soft Matter Research, Freiburg Institute for Advanced Studies,
University of Freiburg, Albertstr. 19, 79104 Freiburg, Germany}

\date{\today}% It is always \today, today,
             %  but any date may be explicitly specified

\begin{abstract}
Critical comments on the entropic lattice Boltzmann equation (ELBE), by Li-Shi Luo, Wei Liao, Xingwang Chen, Yan Peng and Wei Zhang in Ref.\ \cite{Luo2011}, are based on simulations which make use of a model that, despite being called
ELBE by the authors, is in fact fully equivalent to the standard lattice Bhatnagar-Gross-Krook equation.
As a result, the conclusion of  \cite{Luo2011} on ELBE is circular, hence devoid of scientific bearing.
\end{abstract}
\pacs{47.11.-j,~05.20.Dd}

\maketitle

In a recent paper, authors
Li-Shi Luo, Wei Liao, Xingwang Chen, Yan Peng and Wei Zhang, Phys. Rev. E 83, 056710 (2011), claim that that the entropic lattice Boltzmann equation (ELBE)
{\it "does not improve the numerical stability of SRT (single-relaxation-time) or Lattice Bhatnagar-Gross-Krook (LBGK) model".}
They go on by quite categorically stating that
{\it "the ELBE scheme is the most inferior among the LB models tested in the study, thus is
unfit for carrying out numerical simulations in practice".}

In this Comment, we point out that the above statements by  \cite{Luo2011} do not
bear scientific relevance. The reason is simple: what Luo et al
implemented as ELBE, is  not ELBE, but LBGK in disguise.

Since the correct description of ELBE was not presented in \cite{Luo2011}, let us remind that in the ELBE scheme, populations associated with  the discrete velocities $\bm{v}_i$ evolve according to the following kinetic equation,
%(the index $i$ runs over the set of discrete velocities)

\begin{equation}\label{eq:ELBGK}
f_i(\bm{x}+\bm{v}_i,t+1)-f_i(\bm{x},t)=\alpha\beta(\feq_i-f_i),
\end{equation}
In the above, $\feq_i$ is the local equilibrium, which minimizes the entropy function $H=\sum_i f_i\ln(f_{i}/W_{i})$,
where weights $W_i$ are lattice-specific.
In Eq.\ (\ref{eq:ELBGK}), $\alpha$ is the maximal over-relaxation parameter, which is
operationally available as the positive root of the entropy condition (absence of subscripts denotes the
full set of discrete populations):
\begin{equation}\label{eq:Hestimate}
H(\alpha\feq+(1-\alpha)f)-H(f)=0.
\end{equation}
This entropy estimate is key, as it assures the discrete-time $H$-theorem: For $\beta\in[0,1]$, the total entropy $\bar{H}(t)=\sum_{\bm{x}}H(f(\bm{x},t))$ is not increasing, $\bar{H}(t+1)\le \bar{H}(t)$.
Note that the validity of the $H$-theorem requires not just the equilibrium to be evaluated
through the minimization of $H$ but also, and most importantly,  the fulfillment of  the entropy condition (\ref{eq:Hestimate}).
Finally, whenever the simulation is fully resolved (populations stay close to the local equilibrium), the maximal over-relaxation parameter $\alpha$ becomes fixed automatically to the value $\alpha=2$ \cite{AK3}, and the ELBE (\ref{eq:ELBGK}), (\ref{eq:Hestimate}) self-consistently becomes equivalent to the LBGK equation and
recovers Navier-Stokes equations with the kinematic viscosity $\nu=\cs^2\left(\frac{1}{2\beta}-\frac{1}{2}\right)$,
where $\cs$ is speed of sound (a $O(1)$ lattice-dependent constant).

Instead of comparing with ELBE (\ref{eq:ELBGK}), which, by definition, implements
a self-adjusted relaxation time through the entropy estimate (\ref{eq:Hestimate}),
calculations by Luo et al make use of a {\it constant} relaxation time $\tau$, that is

\begin{equation}\label{eq:LBGK}
f_i(\bm{x}+\bm{v}_i,t+1)-f_i(\bm{x},t)=\frac{1}{\tau}(\feq_i-f_i).
\end{equation}

The only remaining input from ELBE in the above, is the local equilibrium $\feq_i$, given by equation (14) in \cite{Luo2011}.
At this point, however, Eq.\ (\ref{eq:LBGK}) is no longer ELBE, as it is in fact to all
effects and purposes {\it equivalent} to the standard LBGK scheme.

Indeed, the authors correctly admit that {\it "one difference between the ELBE and MRT-LBE is the $O(u^3)$ terms in the odd-order equilibrium moments,"}  and {\it "... the difference in the even-order equilibrium moments ... is of the terms of the order $O(u^4)$."}

This implies that the difference between the entropic equilibrium and its standard polynomial approximation, is on the order of the overall errors of the lattice Boltzmann method for the low Mach number flows, and therefore Eq.\ (\ref{eq:LBGK}) is {\it equivalent} to the standard LBGK method.
Since the $H$-theorem is no longer valid with (\ref{eq:LBGK}),
the authors could have opted to replace the equilibrium by a polynomial
approximation, to at least second order, thereby completing in full the identity between the
standard LBGK and (\ref{eq:LBGK}).

With this assessment, and fulfilling the usual low Mach number restrictions, which is roughly
the case for their simulations, the authors cannot make a "comparison of ELBE with LBGK",
as the difference between (\ref{eq:LBGK}) and LBGK is of the overall order of errors of the LB method for low Mach-number simulations.
Therefore, results described as "ELBE" in \cite{Luo2011}, should have been labeled as "LBGK".

Instead, the authors continue as follows:
{\it "...based on our experience and understanding of
the LBE, it is unclear theoretically how the ELBE with a
constant relaxation parameter $\tau$ can improve the numerical
stability of the LBGK scheme, as it has been advocated [3,4]."}

The reader can easily verify that neither in [3] nor [4] (Refs.\ \cite{Ansumali03} and \cite{RMP} here), there
is any claim about constant relaxation parameter improving stability, and whenever stability of ELBE
was discussed in these papers, the entropy estimate (\ref{eq:Hestimate}) has always been provided
(equation (10) in \cite{Ansumali03} and equation (37) in \cite{RMP}).
All statements about ELBE in \cite{Luo2011} are simply incorrect.

Moreover, other independent authors, who have implemented
the ELBE scheme, have made explicit use of (\ref{eq:Hestimate}) \cite{Keating2007,Spasov2009,Geerdink2009,Yasuda2011}.
It is hard to understand how/why Luo et al could have been
misreading the papers \cite{Ansumali03,RMP}, to the point of attributing wrong statements to them.

It is also informing to note that LBGK and "ELBE" (\ref{eq:LBGK})
perform so similarly that
Luo et al come to the point of writing that
{\it "they are so similar to each other"} that {\it "only the results obtained by using ELBE
are shown in Fig. 4"} (\cite{Luo2011}, page 6).
This twin-behavior alone should have warned the authors, that the two allegedly
different methods were basically the same.

On the other hand, it is clear that Luo et al. do know
what ELBE is, as revealed at the very end of the paper.
On page 23, the authors inform the reader that:
{\it "We did not test the ELBE with a variable relaxation time [52-54], which is supposed
to guarantee numerical stability, because it is computationally
inefficient and unphysical with a viscosity depending on space and time; a stable
but inaccurate, unphysical and inefficient scheme is simply not a viable one."}
Thus, by their own admission,  what Luo et al have implemented is
not ELBE, but just LBGK in disguise.

It is not the scope of this Comment to discuss in detail the effect of the variable relaxation time in the genuine
ELBE scheme, leading to the effective viscosity {\it in under-resolved simulations}.
Still, a few clarifications are in order.
The variable viscosity in ELBE is a built-in sub-grid viscosity.
In the resolved direct numerical simulation, the ELBE viscosity remains constant automatically, and
corresponds to $\alpha=2$, as it has been already mentioned.
When the grid is coarsened, local instabilities due to lack of resolution typically lead to a collapse of LBGK.
It is in this situation that ELBE proceeds with the sub-grid viscosity.
To this regard, we wish to point out that there has never been any mystery as to
the fact that ELBE is a natural extension of LBGK into sub-grid simulations; with the distinctive trait
that the stabilization mechanism is directly informed through the second principle ($H$-theorem).
This is well reflected by the specific way ELBE mends instabilities; most of the time during the simulation
the relaxation parameter remains constant everywhere, so that indeed ELBE collapses to LBGK.
It is only at  the onset of a local instability, that ELBE deploys its built-in entropic
stabilization capability. These stabilization events maybe rare in time and very localized in space
(so are incipient instabilities) but they make the whole difference.
This self-adaptive stabilization shows an elegance and universality (no fine tuning of parameters)
which is simply unknown to any other LB method.
It should be stressed that ELBE is not intended for fully resolved direct numerical simulation (DNS) (in this regime ELBE is identical to LBGK) but rather to extend LBGK to much higher Reynolds numbers, beyond the strict DNS regime.
Although the test case chosen in \cite{Luo2011}
(laminar flow in a two-dimensional lid-driven cavity) is quite standard, on the basis of this test alone, it is
difficult to judge any method for the purpose of DNS.
Quantitative information on the adaptive feature of ELBE in high Reynolds
number flow simulations, can be found in \cite{Keating2007,squarecy}.

The authors of \cite{Luo2011} should know that most, if not all, under-resolved simulations, and sometimes
even slightly over-resolved ones, imply some form of effective viscosity, which {\it is} "dependent on space and time"
(the popular flux-corrected-transport techniques and spectral hyperviscosity being two examples in point, respectively).
A bold and categoric rejection of a "variable viscosity" as "unphysical", just reveals a
very ad-hoc view of computational fluid dynamics.

As to computational inefficiency, again, there has never been any mystery on the fact
that the implementation of the entropic estimate requires the solution of a single
non-linear equation for the running parameter $\alpha$, at each lattice site and time-step.
The corresponding computational overhead, however, is more than
compensated by the much reduced grid demand at high Reynolds numbers \cite{squarecy}.

Summarizing,
ELBE with a constant relaxation time is not ELBE, but basically LBGK in disguise.
Thus, what Luo et al achieve in the end, is
a circular result, i.e. cross-compare minor LBGK variants.
%, i.e. basically LBGK with itself.
In the face of the  lack of supporting evidence, in no less than
25 PRE pages, categoric and over-restrictive statements such as
{\it "the ELBE scheme is the most inferior among the LB models tested in the study..."}
appear to be totally unjustified and devoid of scientific relevance.
Ref.\ \cite{Luo2011} is simply a comparison of MRT, TRT and SRT models in a standard flow (similar to, e.\ g.\ \cite{Wu2004}).

%\bibliography{commentrefs}

\end{document}